\documentclass{article}
\usepackage[utf8]{inputenc}

\usepackage{tikz}
\definecolor{colorSFROLAIC}{HTML}{cc0000}
\definecolor{colorSFROLBIC}{HTML}{f1c232}
\definecolor{colorROL}{HTML}{6aa84f}
\definecolor{colorPROL}{HTML}{3d85c6}

\usepackage[unicode,linktoc=all,hidelinks]{hyperref}
\usepackage[a4paper, margin=2cm]{geometry}
\usepackage{mathtools}
\usepackage{subcaption}

\usepackage{amsmath}
\usepackage{amssymb}
\usepackage{amsthm}
\usepackage{bm}
\usepackage{bbm}
\DeclareMathOperator*{\argmax}{arg\,max}

\DeclareSymbolFont{cmletters}{OT1}{cmr}{m}{n}
\DeclareMathSymbol{\Ups}{\mathalpha}{cmletters}{"7}

\usepackage{verbatim}

\usepackage{diagbox}
\usepackage{float} 
\usepackage{adjustbox}
\usepackage{threeparttable}
\usepackage{booktabs}

\usepackage{algorithm}
\usepackage{algorithmic}

\usepackage[english]{babel}
\usepackage{natbib}

\usepackage{graphicx}

\usepackage[title]{appendix}

\newtheorem{assumption}{Assumption}

\newtheorem{proposition}{Proposition}

\usepackage{algorithm}
\usepackage{algorithmic}

\title{Hierarchical Causal Structure Learning}
\author{Sjoerd Hermes$^{1,2}$, Joost van Heerwaarden$^{1,2}$, Fred van Eeuwijk$^{1}$, Pariya Behrouzi$^{1}$}
\date{%
    $^1$ Mathematical and Statistical Methods, Wageningen University\\%
    $^2$ Plant Production Systems, Wageningen University\\
}

\begin{document}

\maketitle
\begin{abstract}
\noindent Traditional statistical approaches primarily model associations between variables, however many scientific and practical questions require causal methods instead. These methods typically rely on assumptions about an underlying structure and are often represented by a Directed Acyclic Graph (DAG). While causal structures can be learned for single-level data, hierarchical or multi-level settings, where units (e.g., plants, students or patients) are nested within groups (e.g., environments, schools or hospitals), lack suitable methods. This article addresses such settings. These multi-level structures frequently arise in fields such as agriculture, where plants grow within different environments. Building on nonlinear structural causal models, or additive noise models, we propose an approach that facilitates causal structure learning for hierarchical causal models with additive unobserved group-level effects and group-specific causal functions. We also propose a simulation-based manner to compute hard interventions for the estimated causal model. In a simulation study, we show that the proposed method is able to identify the underlying hierarchical causal mechanism. A winter-wheat example is used to showcase how the proposed method can be used in practice and how to interpret its results.

\end{abstract}

\noindent%

\section{Introduction}
Understanding phenomena such as the interaction between plant traits, genotypes, and growing conditions requires more than a description of associations. Structural Causal Models (SCMs) represent causal mechanisms by structural equations and a directed acyclic graph (DAG) (Peters et al., \citeyear{peters2014causal}; \citeyear{peters2017elements}). They thereby provide a language for questions about causal direction and intervention.

Many real-world phenomena have an additional complication: their observations are nested. Plants grow within environments, students attend schools, and patients are treated within hospitals. Units in the same group share measured conditions and may also share unmeasured causes. Treating all unit-level observations as independent observations from one homogeneous population can therefore mix within-group and between-group variation, precisely the distinction that matters for causal interpretation.

Hierarchical causal models provide graphical and structural frameworks for such nested systems (Jensen et al., \citeyear{jensen2020object}; Witty et al., \citeyear{witty2020causal}; Weinstein and Blei, \citeyear{weinstein2024hierarchical}). These frameworks are broader than the model studied here. Our aim is deliberately more specific: causal structure learning in a hierarchical model with observed group-level variables, unobserved group-level variables, observed unit-level variables. Such a hierarchical model contains causal relations between group-level variables, between unit-level variables and from group-level variables to unit-level variables. Examples of the latter are an environmental measurement affecting a plant trait or a school characteristic affecting a student outcome. 

Causal discovery for nonlinear, non-hierarchical SCMs can be done in various ways. One way is to make use of a class of methods that consist of a combination of running regressions and independence tests (Mooij et al., \citeyear{mooij2009regression}; Peters et al., \citeyear{peters2014causal}; Mooij et al., \citeyear{mooij2016distinguishing}). Whilst these methods can accurately recover the underlying causal DAG, they are oftentimes computationally inefficient due to the required independence tests. Another class of methods are score-based methods (Bühlmann et al., \citeyear{buhlmann2014cam}; Nowzohour \& Bühlmann, \citeyear{nowzohour2016score}), which are typically more computationally efficient. Given that hierarchical systems may consist of a large number of variables (as there are two levels of variables), we use a score-based method to learn both the group-level and unit-level causal structures. This score based method is the Causal Additive Model (CAM) method by Bühlmann et al., (\citeyear{buhlmann2014cam}), which assumes additive relationships between a variable and its causes. CAM searches for a causal order, estimates additive regressions, and removes edges that are not supported by the data (B\"uhlmann et al., \citeyear{buhlmann2014cam}). In the proposed method, CAM is used to learn the causal structure among the unit-level variables. The hierarchical structure makes this possible by considering one group at a time. Within a fixed group, all group-level quantities are constant across its units. Their additive contributions can therefore be absorbed into the intercept of each unit-level equation. The remaining variation is described by a Gaussian additive noise model for the unit-level variables, to which CAM can be applied.

The hierarchy also creates limitations. Relations involving a group-level covariate are informed by the number of independent groups, not by the total number of units within or across groups. Moreover, an observed group-level effect cannot be separated from unobserved group-level heterogeneity unless a conditional mean-zero condition is imposed. 

We allow the shape and magnitude of unit-level causal relations to differ by group. To make the terms interpretable, a group-specific function is written as a common function plus a group deviation. 

Finally, once a causal graph and its structural functions have been estimated, hard interventions can be simulated by replacing the intervened equation and evaluating the remaining equations in causal order. 

This work contributes to the growing body of literature on tools for causal discovery on observational data. Our hierarchical causal discovery framework is general and applicable across a wide range of nested real-life phenomena, offering a new lens through which to model and interpret these complex datasets. The novelties proposed in this article are critical for, amongst others, modern agricultural research, where trials increasingly consist of different environments or genotypes (van Eeuwijk et al., \citeyear{van2005statistical}; Hermes et al., \citeyear{hermes2024copula}). To illustrate the usefulness of the proposed framework, we provide an application on hierarchical winter wheat data to discover the causal determinants of yield. 

The remainder of the article is structured as follows: Section \ref{Hierarchical Structural Causal Model} starts with a description of the HSCM model and its extension to obtain the proposed method. Causal structure learning is discussed in Section \ref{Causal structure learning}. Computing hard interventions is discussed in Section \ref{Interventions}. Section \ref{Simulation study} evaluates the performance of the proposed method in terms of causal structure learning. Section \ref{Application on winter wheat data} provides an application on an agricultural dataset pertaining to winter wheat. Finally, Section \ref{Conclusion} concludes the article and discusses potential future extensions.

\section{Hierarchical Structural Causal Model} \label{Hierarchical Structural Causal Model}
We consider $m$ groups indexed by $j=1,\ldots,m$ and $n_j$ units in group $j$, indexed by $i=1,\ldots,n_j$. The notation keeps the two levels visible: $\bm Z_j=(Z_j^1,\ldots,Z_j^q)$ contains the observed variables that have one value for group $j$, whereas $\bm X_{ij}=(X_{ij}^1,\ldots,X_{ij}^p)$ contains the observed variables for unit $i$ in that group. For example, $Z_j^1$ might be the average temperature of environment $j$ and $X_{ij}^1$ the yield of plot $i$ in that environment.

\begin{figure}[H]
\centering
\includegraphics[width=0.4\textwidth]{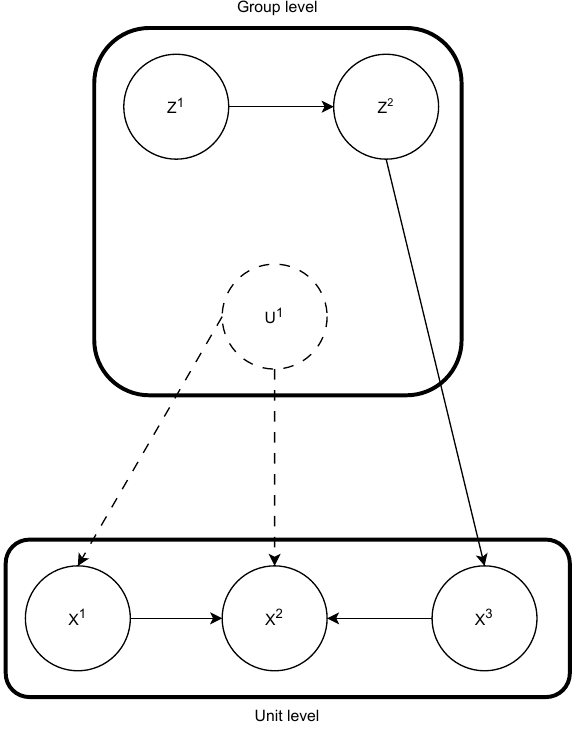}
\caption{Example of a DAG $D_{ZX}$ with two observed group-level variables, one unobserved group-level variable, and three observed unit-level variables.}
\label{fig:ex_graph}
\end{figure}

Unobserved group-level variables are denoted by $\bm U$. These variables may affect one or more unit-level variables, but they do not affect the observed group-level variables $\bm Z$. We write $D_Z$ for the observed group-level DAG, $D_X$ for the causal DAG among the observed unit-level variables, and $D_{ZX}$ for the graph obtained by combining $D_Z$, $D_X$, and the observed arrows from $\bm Z$ to $\bm X$. For a unit-level variable $X^k$, $\mathrm{Pa}_Z(X^k)$ and $\mathrm{Pa}_X(X^k)$ denote its observed group-level and observed unit-level parents. Arrows between the observed levels may point from $\bm Z$ to $\bm X$, but not from $\bm X$ to $\bm Z$. An example of the type of DAG $D_{ZX}$ considered here is shown in Figure \ref{fig:ex_graph}. 

\subsection{Baseline model}
The baseline version of the HSCM is introduced first. At the group level, the baseline model is,
\begin{equation}
Z_j^k = \mu_{Z,k}+\sum_{Z^\ell\in\mathrm{Pa}(Z^k)} f_{k\ell}(Z_j^\ell)+\eta_j^k,
\qquad k=1,\ldots,q.
\label{eq:group_sem}
\end{equation}
At the unit level, the baseline model is
\begin{equation}
X_{ij}^k
= \mu_{X,k}
+\sum_{Z^\ell\in\mathrm{Pa}_Z(X^k)}\beta_{k\ell}Z_j^\ell
+A_j^k
+\sum_{X^d\in\mathrm{Pa}_X(X^k)} s_{kd}(X_{ij}^d)
+\varepsilon_{ij}^k,
\qquad k=1,\ldots,p.
\label{eq:HSCM_detailed}
\end{equation}
The terms have direct interpretations. The coefficient $\beta_{k\ell}$ describes a linear direct effect of the group-level variable $Z^\ell$ on $X^k$. The smooth function $s_{kd}$ describes the possibly nonlinear direct effect of a unit-level variable $X^d$ on $X^k$. $\eta_j^k$ and $\varepsilon_{ij}^k$ are noise variables that vary between groups and units respectively. Finally, $A_j^k$ collects additive contributions to $X^k$ that are shared by all units in group $j$ but are not explained by the observed $\bm Z_j$. For example, if an unobserved group-level variable $U_j^r$ enters the equation through $h_{kr}(U_j^r)$, that contribution is included in $A_j^k$. Components of $\bm A_j=(A_j^1,\ldots,A_j^p)$ may be dependent, so one latent group-level factor may influence several unit-level variables. The following assumptions are needed to determine the causal structure and effects from the observed data. 


\begin{assumption}[Independent groups and group-level noise]\label{ass:group_sampling}
The groups are independent realizations of the same
data-generating process. The observed group-level variables $\bm Z_j$
follow the causal model in Equation (\ref{eq:group_sem}). The group-level
noise vectors $\bm\eta_j$ are independent across groups, and within each
group the noise variables associated with different group-level variables
are mutually independent, with
\begin{equation}
\eta_j^k\sim\mathcal N(0,\sigma_{Z^k}^2),
\qquad
\sigma_{Z^k}^2>0,
\qquad
k=1,\ldots,q.
\label{eq:group_noise_gaussian}
\end{equation}
\end{assumption}
\noindent This assumption ensures that groups provide independent information and
that the group-level noise variables satisfy the Gaussian additive noise
model used by CAM. 

\begin{assumption}[Independent units and unit-level noise]\label{ass:sampling}
Conditional on $\bm Z_j$, $\bm A_j$, and the group-specific functions, the units within group $j$ are independent and generated by the same unit-level additive noise model. The noise vectors $\bm\varepsilon_{ij}$ are independent across units, and within each unit the noise variables associated with different unit-level variables are mutually independent, with
\begin{equation}
\varepsilon_{ij}^k\sim\mathcal N(0,\sigma_{X^k}^2),
\qquad
\sigma_{X^k}^2>0,
\qquad
k=1,\ldots,p.
\label{eq:unit_noise_gaussian}
\end{equation}
\noindent The noise distribution is the same across groups. Therefore,
\begin{equation}
E(\varepsilon_{ij}^k\mid \bm Z_j,\bm A_j)=0
\qquad\text{for every }k,
\label{eq:unit_noise_mean_zero}
\end{equation}
\noindent which also implies
\begin{equation*}
E(\varepsilon_{ij}^k\mid\bm Z_j)=0.
\end{equation*}
\end{assumption}

\noindent This assumption ensures that the units within a group provide independent information for learning the unit-level structure. The independent Gaussian noise variables are consistent with CAM, while the conditional mean-zero property is also used when identifying group-to-unit effects. The residual variance for each unit-level variable is assumed to be the same across groups.

\begin{assumption}[Conditional mean-zero assumption for unobserved group effects]\label{ass:group_conditional_mean_zero}

For every unit-level equation $k$,

\begin{equation}
E(A_j^k\mid \bm Z_j)=0.
\label{eq:group_conditional_mean_zero}
\end{equation}
\end{assumption}

\noindent This assumption allows the effect of an observed group-level variable on $X^k$ to be separated from unobserved between-group variation. If it does not hold, the estimated group-to-unit effect may combine the causal effect of $\bm Z$ with unobserved group-level effects.

\begin{assumption}[Centering of additive components]\label{ass:additive_centering}

Additive functions are centered with respect to their corresponding marginal distributions. For every group-level edge $Z^\ell\rightarrow Z^k$ and unit-level edge $X^d\rightarrow X^k$,

\begin{equation}
E_{P_{Z^\ell}}\{f_{k\ell}(Z^\ell)\}=0
\quad\text{and}\quad
E_{P_{X^d}}\{s_{kd}(X^d)\}=0,
\label{eq:cam_centering}
\end{equation}

\noindent where $P_{Z^\ell}$ denotes the marginal group-level distribution of $Z^\ell$, and $P_{X^d}$ denotes the marginal distribution across units, taking the group structure into account.
\end{assumption}
\noindent This assumption is needed, as additive components are defined only up to constants unless a normalization is imposed (Bühlmann et al., \citeyear{buhlmann2014cam}).

\begin{assumption}[Shared observed unit-level DAG]\label{ass:shared_dx}
There is a single DAG $D_X$ on $X^1,\ldots,X^p$ such that the presence and orientation of every direct observed unit-level edge are the same in all groups. Causal functions may differ by group, but for every edge in $D_X$ the group-specific structural function is nonconstant on the relevant support in every group.
\end{assumption}
\noindent This assumption allows the unit-level DAG $D_X$ to be learned from the data of a single group. Considering using a single group avoids between-group variation entering the estimation of the unit-level causal structure. 

\begin{assumption}[Conditions for CAM structure learning]\label{ass:cam_regular}
Except for unobserved group-level variables that affect unit-level variables, no other unobserved confounders exist. All variables have a nondegenerate joint distribution. The noise variables are mutually independent, Gaussian, and have positive variances, as specified in Assumptions \ref{ass:group_sampling} and \ref{ass:sampling}. All causal group-group and unit-unit funtions are nonlinear and three times differentiable, and the remaining conditions required for CAM identification in B\"uhlmann et al. (\citeyear{buhlmann2014cam}) are assumed to hold.
\end{assumption}

\noindent These conditions allow us to apply the existing CAM identification result rather than derive it again here. 

\subsection{Group-specific unit-level functions}
\label{sec:group_specific_functions}

It may not be realistic to assume that the causal functions between unit-level variables are identical across groups. We therefore allow the causal function associated with an edge $X^d\rightarrow X^k$ to differ by group. Let $s_{kd}^{(j)}$ denote the causal function for this edge in group $j$. In Equation (\ref{eq:HSCM_detailed}), each $s_{kd}$ is then replaced by $s_{kd}^{(j)}$.

To compare these group-specific functions, they must first be considered over a common range of $X^d$. Because the distribution of $X^d$ may differ between groups, define
\begin{equation}
\mathcal S_d = \bigcap_{j=1}^m \operatorname{supp}(X^d\mid J=j),
\label{eq:common_support}
\end{equation}
where $\mathcal S_d$ denotes the common support of $X^d$ across groups.

The group-specific functions must also use the same centering convention. Otherwise, an arbitrary constant could be added to one function and subtracted from the group-specific intercept, making comparisons between groups ambiguous. Let $\nu_d$ be a fixed distribution on $\mathcal S_d$ that is used as a common reference for centering.

\begin{assumption}[Centering of group-specific causal functions]
\label{ass:centered_deviations}
For every unit-level edge $X^d\rightarrow X^k$, the group-specific causal
functions are centered using the same reference distribution:
\begin{equation}
E_{\nu_d}\{s_{kd}^{(j)}(X^d)\}=0, \qquad \text{for every }j.
\label{eq:group_function_centering}
\end{equation}
\end{assumption}

\noindent Using the same reference distribution ensures that the group-specific functions are centered in the same way and can therefore be compared directly across groups.

We next separate each group-specific function into a common part and a group-specific deviation. To define the common part, let $w_j>0$ be prespecified group weights satisfying

\begin{equation*}
\sum_{j=1}^m w_j=1.
\end{equation*}

\noindent
The weights determine how much each observed group contributes to the
common function. With equal weights, $w_j=1/m$, every group contributes
equally. We define the common causal function as the weighted average

\begin{equation}
s_{kd}(x) = \sum_{j=1}^m w_j s_{kd}^{(j)}(x), \qquad x\in\mathcal S_d.
\label{eq:common_function}
\end{equation}

\noindent
The deviation for group $j$ is then defined by
\begin{equation}
\delta_{j,kd}(x) = s_{kd}^{(j)}(x)-s_{kd}(x).
\label{eq:group_deviation}
\end{equation}
\noindent Hence, each group-specific causal function can be written as

\begin{equation}
s_{kd}^{(j)}(x) = s_{kd}(x)+\delta_{j,kd}(x).
\label{eq:group_specific_decomp}
\end{equation}

\noindent Here, $s_{kd}$ represents the part of the causal relation that is common across the observed groups, while $\delta_{j,kd}$ describes how the causal function in group $j$ differs from this common function.

The definitions above have two useful consequences. First, because every group-specific function is centered using the same distribution $\nu_d$,

\begin{equation*}
E_{\nu_d}\{s_{kd}(X^d)\}=0, \qquad E_{\nu_d}\{\delta_{j,kd}(X^d)\}=0.
\end{equation*}

\noindent Second, the group-specific deviations average to zero according to the chosen group weights:

\begin{equation*}
\sum_{j=1}^m w_j\delta_{j,kd}(x)=0, \qquad \text{for every }x\in\mathcal S_d.
\end{equation*}

\noindent The common centering convention and weighted-average definition therefore provide a unique way to separate the common function from the group-specific deviations. 

The centering convention is also important when estimating the unobserved group effect $A_j^k$. Because each group-specific causal function is centered, constant shifts are absorbed by $A_j^k$ rather than by $s_{kd}^{(j)}$. This preserves the interpretation of Assumption \ref{ass:group_conditional_mean_zero}.


\section{Causal structure learning} \label{Causal structure learning}
The observed graph is estimated in layers. The group-level graph $D_Z$ is learned from one observation per group, the observed unit-level graph $D_X$ is learned from all units in a single group, and observed cross-level arrows $\bm Z\rightarrow\bm X$ are learned after the unit-level parent sets have been estimated. 

\subsection{How CAM works}\label{sec:cam_explanation}
CAM is the main causal-discovery engine used in the proposed method. It is therefore useful to distinguish clearly what CAM assumes, how CAM works, and which parts of the HSCM are learned by CAM.

For a collection of $r$ observed variables $\bm V=(V^1,\ldots,V^r)$, CAM assumes an additive structural equation model of the form
\begin{equation}
V^k=\mu_k+\sum_{V^\ell\in\mathrm{Pa}(V^k)}f_{k\ell}(V^\ell)+e^k,
\qquad k=1,\ldots,r,
\label{eq:cam_generic}
\end{equation}
where the graph is acyclic, all variables relevant to the graph are observed, and the noise variables are mutually independent with
\begin{equation}
e^k\sim\mathcal N(0,\sigma_k^2),
\qquad \sigma_k^2>0.
\label{eq:cam_gaussian_noise}
\end{equation}
Each parent contributes through its own univariate function. The functions are centred, $E\{f_{k\ell}(V^\ell)\}=0$, so that their constants are absorbed by $\mu_k$. Under the nonlinear smooth-function and regularity conditions summarized in Assumption \ref{ass:cam_regular}, the observational distribution identifies the DAG rather than only its Markov equivalence class (Bühlmann et al., \citeyear{buhlmann2014cam}). 

CAM distinguishes between two different tasks. First, it estimates a causal order, that is, an ordering of the variables in which causes appear before their effects. It then determines which of the possible edges that are consistent with this order are actually present. Let $\pi$ denote an ordering of the $r$ variables, so that $V^{\pi(1)},\ldots,V^{\pi(r)}$ are arranged from first to last in that order. If this causal order were known, every parent of $V^{\pi(a)}$ would have to occur among the preceding variables $V^{\pi(1)},\ldots,V^{\pi(a-1)}$.. For a proposed order, CAM therefore fits an additive regression of each variable on its predecessors and computes the residual standard deviation $\widehat\sigma_{\pi(a),\pi}$. Apart from constants, the Gaussian negative log-likelihood is represented by the score
\begin{equation}
\widehat S(\pi)=\sum_{a=1}^r\log\widehat\sigma_{\pi(a),\pi},
\label{eq:cam_order_score}
\end{equation}
so smaller values indicate an order whose additive regressions fit better. Under Equation (\ref{eq:cam_gaussian_noise}), this is the correctly specified Gaussian likelihood score. 

A DAG may be compatible with more than one causal order. CAM therefore only needs to find an order that is consistent with the true DAG; the final edges are determined during pruning. The CAM algorithm consists of three stages:

\begin{enumerate}

\item \textbf{Preliminary neighbourhood selection (PNS; optional).}
PNS reduces the number of candidate parents considered for each variable. It does not determine edge directions or the final graph. If a true parent is excluded at this stage, that edge cannot be recovered later. HSCM does not use this step.

\item \textbf{Order search (IncEdge).}
Because evaluating all possible orders is usually infeasible, CAM uses a greedy search. Starting from an empty graph, it repeatedly adds the edge that most improves the likelihood while avoiding directed cycles. The resulting dense acyclic graph represents an estimated causal order rather than the final DAG.

\item \textbf{Pruning and function estimation (Prune).}
Given the estimated order, each variable is regressed on its possible parents using an additive model. Edges that are not supported by the data are removed. Pruning can remove edges but cannot reverse them, since their directions are already determined by the estimated order. The remaining edges define the estimated DAG, and the fitted smooth functions provide estimates of the corresponding causal functions. 

\end{enumerate}
\noindent Within the HSCM, CAM is used in two specific places. First, if the number of groups is adequate and the group-level model is scientifically relevant, CAM is applied to the $m$ group-level observations in $\bm Z$ (one observation per group) to estimate $D_Z$. Second, CAM is applied to observations from one group to estimate the shared $D_X$. CAM does not estimate the cross-level arrows $\bm Z\rightarrow\bm X$: their direction is fixed by the hierarchical model, and their coefficients are estimated using the grouped regression described in Section \ref{sec:cross_level_learning}. 

Some details on the estimation of $D_X$ and the $\bm Z\rightarrow\bm X$ relations are provided in the following subsections. No additional details are provided on the estimation of $D_Z$, as the application of CAM on the group-level observations is straightforward.

\subsection{Using one group to estimate $D_X$}
To show that $D_X$ can be estimated by using the data from one group, we write
 \[
c_j^k=\mu_{X,k}
+\sum_{Z^\ell\in\mathrm{Pa}_Z(X^k)}\beta_{k\ell}Z_j^\ell
+A_j^k.
\]
Because $j$ is fixed, $c_j^k$ is a constant rather than a variable across units. Allowing the group-specific functions from Equation (\ref{eq:group_specific_decomp}), the unit-level equation becomes
\begin{equation}
X_{ij}^k=c_j^k+
\sum_{X^d\in\mathrm{Pa}_X(X^k)}s_{kd}^{(j)}(X_{ij}^d)
+\varepsilon_{ij}^k,
\label{eq:within_group_cam}
\end{equation}
which has exactly the Gaussian additive form required by CAM. Pooling groups would make $c_j^k$ vary with $j$ and would generally reintroduce the between-group variation that this reduction removes. Thus, we can formally make the following statement.

\begin{proposition}[Reduction to a Gaussian additive noise model within a group]\label{prop:within_group}
Under Equation (\ref{eq:HSCM_detailed}), allowing the group-specific functions in Equation (\ref{eq:group_specific_decomp}), and under Assumptions \ref{ass:sampling}, \ref{ass:shared_dx}, and \ref{ass:cam_regular}, fix a group $j$. Within that group, the observed unit-level equations form a Gaussian additive noise model with graph $D_X$, equation-specific intercepts, functions $s_{kd}^{(j)}$, and mutually independent noises $\varepsilon_{ij}^k\sim\mathcal N(0,\sigma_{X,k}^2)$. Hence $D_X$ is identifiable from the within-group observational distribution under the CAM identifiability conditions.
\end{proposition}
\noindent A proof is given in the Appendix.

Any group whose conditional distribution satisfies the CAM assumptions and has adequate support can identify the shared $D_X$ in the population. In practice, the group with the largest $n_j$ is used to estimated $D_X$.

\subsection{Group-to-unit relations}
\label{sec:cross_level_learning}

After estimating the unit-level DAG $D_X$, we next determine which group-level variables directly affect each unit-level variable. These group-to-unit relations are learned from variation between groups, since the group-level variables $\bm Z_j$ are constant for all units within group $j$.

Consider the equation for a unit-level variable $X^k$. To show how the group-to-unit effects can be identified, suppose for the moment that its unit-level parent set and corresponding causal functions are known. Define
\begin{equation} 
Q_{ij}^k = X_{ij}^k - \sum_{X^d\in\mathrm{Pa}_X(X^k)} s_{kd}^{(j)}(X_{ij}^d),
\label{eq:crosslevel_residual}
\end{equation}

\noindent so that $Q_{ij}^k$ is $X_{ij}^k$ after removing the contributions of its unit-level parents. For the baseline model in Equation (\ref{eq:HSCM_detailed}), $s_{kd}^{(j)}=s_{kd}$. Substituting the causal equation for $X^k$ gives
\begin{equation}
Q_{ij}^k = \mu_{X,k} + \sum_{Z^\ell\in\mathrm{Pa}_Z(X^k)} \beta_{k\ell}Z_j^\ell + A_j^k + \varepsilon_{ij}^k.
\label{eq:crosslevel_reduced}
\end{equation}

\noindent Thus, after accounting for the unit-level parents, the remaining terms are the observed group-level effects, the unobserved group effect $A_j^k$, and the unit-level noise $\varepsilon_{ij}^k$.

To isolate the observed group-level effects, consider the conditional expectation given $\bm Z_j$. By Assumption \ref{ass:sampling},
\[
E(\varepsilon_{ij}^k\mid\bm Z_j)=0.
\]
\noindent The unobserved group effect $A_j^k$ may also contribute to differences between groups. To separate this variation from the observed group-to-unit effects, Assumption \ref{ass:group_conditional_mean_zero} requires
\[
E(A_j^k\mid\bm Z_j)=0.
\]
\noindent Therefore,
\begin{equation}
E(Q_{ij}^k\mid\bm Z_j) = \mu_{X,k} + \sum_{Z^\ell\in\mathrm{Pa}_Z(X^k)} \beta_{k\ell}Z_j^\ell.
\label{eq:crosslevel_conditional_mean}
\end{equation}

\noindent Hence, the direct group-to-unit effects are identified from how the conditional mean of $Q_{ij}^k$ changes with the observed group-level variables. The following proposition formalizes this result.
\begin{proposition}[Identification of linear group-to-unit effects]
\label{prop:cross_level}

Let $S_k\subseteq\{1,\ldots,q\}$ be a prespecified set of candidate group-level parents of $X^k$ that contains all true observed group-level parents. Let $\bm Z_{j,S_k}$ denote the corresponding subvector of $\bm Z_j$, and set $\beta_{k\ell}=0$ for candidates that are not true parents.

Assume that the unit-level parent set and corresponding causal functions in Equation (\ref{eq:crosslevel_residual}) are identified. Suppose Assumptions \ref{ass:group_sampling}, \ref{ass:sampling}, and \ref{ass:group_conditional_mean_zero} hold, and that the second-moment matrix of
\[
W_{j,k} = (1,\bm Z_{j,S_k}^{\top})^{\top}
\]

\noindent is nonsingular. Then $\mu_{X,k}$ and the group-to-unit coefficients $\bm\beta_{k,S_k}$ are uniquely identified from
\begin{equation}
E(Q_{ij}^k\mid\bm Z_{j,S_k}) = \mu_{X,k} + \bm\beta_{k,S_k}^{\top}\bm Z_{j,S_k}.
\label{eq:crosslevel_moment}
\end{equation}

\end{proposition}

\noindent A proof is given in the Appendix.




\subsection{Estimation algorithm}
Algorithm \ref{alg:alg_structurelearn} summarizes the DAG-learning procedure under the assumptions stated above.

\begin{algorithm}[H]
\caption{HSCM structure learning}
\label{alg:alg_structurelearn}
\textbf{Input:} Data $\bm Z,\bm X$, pruning threshold $\alpha$.\\
\textbf{Output:} Estimated observed graph $\widehat D_{ZX}$, estimated structural causal functions, and estimated equation-specific Gaussian noise variances.
\begin{algorithmic}[1]
\STATE If required and supported by the number of groups, estimate $\widehat D_Z=\mathrm{CAM}(\bm Z)$.
\STATE Set $\tilde j=\argmax_j n_j$ and estimate $\widehat D_X=\mathrm{CAM}(\bm X_{\tilde j})$.
\FOR{$k=1,\ldots,p$}
	\STATE Fit a regression for $X^k$ with the candidate group-level variables as linear effects, the estimated unit-level parents as smooth effects, and a random intercept for group.
    \STATE Retain $Z^\ell\rightarrow X^k$ when the coefficient for $Z^\ell$ satisfies $p\leq\alpha$.
\ENDFOR
\STATE Combine $\widehat D_Z$, $\widehat D_X$, and the retained cross-level arrows.
\STATE Refit structural functions conditional on the assembled graph and estimate the Gaussian noise variance of each fitted equation. For group-specific functions, impose the constraints in Assumption \ref{ass:centered_deviations}.
\end{algorithmic}
\end{algorithm}

\noindent When the group-level DAG $D_Z$ is not scientifically relevant, or when the number of groups is too small to support reliable group-level structure learning, Step 1 can be omitted. The same limitation applies to cross-level regression: a large number of units within only a few groups does not create additional independent information about variables that are constant within group. Following Bühlmann et al.\ (\citeyear{buhlmann2014cam}), we set $\alpha$ to 0.001 in practice.

\section{Interventions} \label{Interventions}
A SCM supports interventions in addition to observational prediction (Pearl, \citeyear{pearl2009causality}; Peters et al., \citeyear{peters2017elements}). This also holds for the HSCM. We consider hard interventions $\mathrm{do}(Y=y)$. Such an intervention replaces the causal equation for $Y$ by the constant assignment $Y=y$.

The fitted HSCM can be used to simulate interventions on observed group-level and unit-level variables across the observed groups. Consider first an intervention on a unit-level variable $X^k$. The same intervention is applied separately to each observed group $j$. Within each group, its observed group-level variables $\bm Z_j$, estimated group effect  $\widehat{\bm A}_j$, and estimated group-specific causal functions are kept fixed. The causal equation for $X^k$ is replaced by the intervention value. New unit-level noise variables are then drawn from their fitted Gaussian distributions, after which the remaining unit-level variables are generated in causal order.

For an intervention on an observed group-level variable $Z^\ell$, the equation for $Z^\ell$ is replaced by the intervention value and the remaining group-level variables are generated in causal order. The resulting group-level values are then propagated through the group-to-unit relations and the unit-level DAG. To ensure that the causal interpretation for group-level interventions, we require another assumption.

\begin{assumption}[Group characteristics under intervention]
\label{ass:intervention_group_characteristics}
For any considered intervention on a group-level variable $Z^\ell$, the unobserved group-level variables contributing to $\bm A_j$, and any unobserved variables determining the group-specific causal functions, are not descendants of $Z^\ell$.
\end{assumption}

\noindent Under this assumption, intervening on $Z^\ell$ does not change the group effect $\bm A_j$ or the group-specific causal functions. These quantities can therefore be kept fixed when simulating the intervention.

In both cases, simulation proceeds in causal order, meaning that the parents of each variable are generated before the variable itself. For unit-level simulation, this requires the fitted unit-level graph, causal functions, group effects, group-specific causal functions, and unit-level noise variances. For an intervention on a group-level variable, the fitted group-level
equations and group-level noise variances are additionally required. Algorithm \ref{alg:alg_intervention} provides an overview of how hard interventions are simulated using the HSCM.

Interventions for new or unseen groups require additional assumptions and are outside the scope of the current procedure.

\begin{algorithm}[H]
\caption{Simulation of a hard intervention for the observed groups}
\label{alg:alg_intervention}
\textbf{Input:} Fitted acyclic HSCM, intervention
$\mathrm{do}(Y=y)$, number of simulation replicates $B$, and number of
simulated units per group $N$.\\
\textbf{Output:} Simulated interventional data.
\begin{algorithmic}[1]
\FOR{$j=1,\ldots,m$}
    \STATE Hold the fitted group effect $\widehat{\bm A}_j$ and the
    fitted group-specific causal functions for group $j$ fixed.
    \FOR{$b=1,\ldots,B$}
        \IF{$Y$ is a group-level variable}
            \STATE Draw mutually independent group-level noise variables
            $\widetilde\eta_{jb}^k
            \sim
            \mathcal N(0,\widehat\sigma_{Z^k}^2),
            \qquad k=1,\ldots,q.$
            \FOR{$Z^k$ in a causal order of $\widehat D_Z$}
                \IF{$Z^k=Y$}
                    \STATE Set $\widetilde Z_{jb}^k=y$.
                \ELSE
                    \STATE Evaluate the fitted causal equation for $Z^k$
                    using its simulated parents and the newly drawn noise
                    $\widetilde\eta_{jb}^k$.
                \ENDIF
            \ENDFOR
        \ELSE
            \STATE Set $\widetilde{\bm Z}_{jb}=\bm Z_j$.
        \ENDIF
        \STATE Draw $N$ independent unit-level noise vectors with
        mutually independent components $\widetilde\varepsilon_{jbi}^k
        \sim
        \mathcal N(0,\widehat\sigma_{X^k}^2)$.
        \FOR{$i=1,\ldots,N$}
            \FOR{$X^k$ in a causal order of $\widehat D_X$}
                \IF{$X^k=Y$}
                    \STATE Set $\widetilde X_{jbi}^k=y$.
                \ELSE
                    \STATE Evaluate the fitted causal equation for $X^k$
                    using its simulated unit-level parents,
                    $\widetilde{\bm Z}_{jb}$,
                    $\widehat A_j^k$, the fitted group-specific causal
                    functions, and
                    $\widetilde\varepsilon_{jbi}^k$.
                \ENDIF
            \ENDFOR
        \ENDFOR
    \ENDFOR
\ENDFOR
\end{algorithmic}
\end{algorithm}

\section{Simulation study} \label{Simulation study}

The effectiveness of the proposed method is demonstrated by means of a simulation study. Two variants of the proposed method are considered: the baseline HSCM without group-specific unit-unit functions, see Equations (\ref{eq:group_sem}) and (\ref{eq:HSCM_detailed}), and the extended HSCM with group-specific unit-unit functions, see Equation (\ref{eq:group_specific_decomp}). First, the causal graph is generated to determine the causal order. The causal graph consists of $q \in \{4,10\}$ observed group-level variables, $R \in \{1,2\}$ unobserved group-level variables and $p \in \{4,10\}$ observed unit-level variables. To generate both the observed group-level and unit-level DAGs, respectively $D_Z$ and $D_X$, either 2 or 5 edges are randomly sampled, depending on whether $p = q = 4$ or $p = q = 10$ respectively. Subsequently, the same number of edges are sampled between any of the group-level and unit-level variables. Finally, either 2 or 4 edges are sampled from the unobserved group-level variables to the unit-level variables, depending on whether 1 or 2 unobserved group-level variables are generated respectively. 

With the full DAG created the data can be simulated. The data consist of $m \in \{25,50,100,250\}$ groups and $n \in \{25,50,100,250\}$ units within groups. All data for the group-level variables are simulated first, starting with the source variables (variables without parents), which only consist of a (standard normal) noise variable. The other group level variables are simulated according to the causal order corresponding with $D_Z$. The nonlinear functions used are the sin, squared, cubic, exponential relu and softplus functions, where is function is randomly assigned a positive or negative sign and a coefficient in $[0.5,2]$. Similarly, the data for the unit-level variables are generated starting with the source variables. Subsequently, the other unit-level variables are generated following the causal order in $D_X$, combined with the group-level parents of the unit-level variables. The same nonlinear functions with a random sign and coefficient that were used for the group-level variables are used to generate the unit-level variables.
\\
\\
The HSCM is fitted on the generated data. Its performance is evaluated by means of the Structural Hamming Distance (SHD) and the root mean squared error (RMSE). The SHD is used as an error metric for the causal structure, that is, how does the estimated DAG $\hat{D}_{ZX}$ compare to the true DAG $D_{ZX}^0$, whereas the RMSE is used as an error metric for the estimated causal functions. The RMSE is computed using only the correctly estimated relations in the DAG $D_{ZX}$. Whilst the RMSE is well-known to the general audience, the SHD is not. The SHD counts the number of operations required to go from the estimated DAG $\hat{D}_{ZX}$ to the true DAG $D_{ZX}^0$. The RMSE for the causal functions is estimated by evaluating the true and estimated functions on the same set of values and computing the RMSE on the respective outcomes. For each simulation setting consisting of different $n,m,p$ and $q$, the model is fitted on 50 generated datasets. To showcase the relative performance of the proposed method, the original CAM method is also fitted on this simulated data, for which the same error metrics are computed. While CAM represents a non-hierarchical baseline, rather than as a competing method specifically designed for hierarchical causal discovery, no such method exists to the best of our knowledge. Results for this simulation study are provided in Table \ref{tab:simres1}. Additional simulation results, for data generated without Gaussian errors, are provided in the Appendix.

\begin{table}[H]
\centering
\scalebox{0.8}{%
 \begin{threeparttable}
  \caption{Simulation results. The Structural Hamming Distance (SHD) and root mean squared error (RMSE) are averaged across 50 randomly generated datasets. The RMSE is computed only on correctly estimated relations in the DAG $D_{ZX}$.}
  \label{tab:simres1}
     \begin{tabular}{c | cc | cc | cc | cc }
        \toprule
        \midrule
         & \multicolumn{4}{c|}{CAM} & \multicolumn{4}{c}{HSCM}\\ \midrule
         & \multicolumn{2}{c|}{No group-specific functions} & \multicolumn{2}{c|}{Group-specific functions} & \multicolumn{2}{c|}{No group-specific functions} & \multicolumn{2}{c}{Group-specific functions}\\ \midrule
         $\bm{n, m, p = q}$ & \textbf{SHD} & \textbf{RMSE} & \textbf{SHD} & \textbf{RMSE} & \textbf{SHD} & \textbf{RMSE} & \textbf{SHD} & \textbf{RMSE}\\ \midrule
$25, 25, 4$ & 11.20 (1.46) & 1.00 (0.78) & 12.00 (1.30) & 0.49 (0.02) & 2.88 (0.19) & 0.12 (0.02) & 2.94 (0.17) & 0.05 (0.01)\\ 
$50, 25, 4$ & 11.40 (1.25) & 1.05 (0.85) & 12.20 (1.71) & 0.49 (0.03) & 2.62 (0.23) & 0.11 (0.02) & 2.52 (0.15) & 0.04 (0.01)\\ 
$100, 25, 4$ & 11.60 (1.29) & 0.86 (0.68) & 13.00 (1.73) & 0.77 (0.27) & 2.20 (0.22) & 0.09 (0.01) & 2.32 (0.16) & 0.03 (0.01)\\ 
$250, 25, 4$ & 11.40 (1.25) & 0.92 (0.74) & 13.00 (1.73) & 0.76 (0.27) & 1.94 (0.20) & 0.07 (0.01) & 2.24 (0.20) & 0.02 (0.03)\\ 
$25, 50, 4$ & 11.40 (1.54) & 0.14 (0.08) & 15.40 (0.68) & 0.35 (0.01) & 2.42 (0.18) & 0.06 (0.01) & 2.44 (0.18) & 0.07 (0.01)\\ 
$50, 50, 4$ & 11.00 (2.05) & 0.50 (0.29) & 17.40 (1.08) & 0.38 (0.02) & 2.18 (0.21) & 0.06 (0.01) & 2.08 (0.17) & 0.09 (0.02)\\ 
$100, 50, 4$ & 11.40 (1.94) & 0.03 (0.01) & 18.60 (1.12) & 0.37 (0.01) & 1.64 (0.19) & 0.06 (0.01) & 1.76 (0.15) & 0.12 (0.02)\\ 
$250, 50, 4$ & 11.40 (1.96) & 0.03 (0.01) & 18.40 (1.08) & 0.37 (0.01) & 1.46 (0.17) & 0.09 (0.01) & 1.94 (0.20) & 0.08 (0.02)\\ 
$25, 100, 4$ & 9.40 (1.21) & 1.05 (0.57) & 13.40 (0.93) & 0.28 (0.02) & 2.26 (0.17) & 0.06 (0.01) & 2.00 (0.14) & 0.03 (0.01)\\ 
$50, 100, 4$ & 10.20 (0.80) & 0.59 (0.39) & 15.40 (1.03) & 0.38 (0.10) & 1.56 (0.15) & 0.06 (0.01) & 1.66 (0.18) & 0.03 (0.01)\\ 
$100, 100, 4$ & 10.40 (1.03) & 0.78 (0.53) & 14.80 (1.11) & 0.30 (0.02) & 1.26 (0.16) & 0.06 (0.01) & 1.40 (0.16) & 0.02 (0.00)\\ 
$250, 100, 4$ & 9.60 (1.50) & 0.70 (0.53) & 14.80 (0.86) & 0.29 (0.02) & 0.74 (0.11) & 0.05 (0.01) & 1.12 (0.16) & 0.03 (0.01)\\ 
$25, 250, 4$ & 9.80 (0.86) & 0.13 (0.04) & 15.20 (1.39) & 0.20 (0.02) & 1.88 (0.16) & 0.05 (0.01) & 1.50 (0.12) & 0.02 (0.00)\\ 
$50, 250, 4$ & 9.00 (1.05) & 0.17 (0.05) & 18.00 (0.84) & 0.20 (0.02) & 1.20 (0.20) & 0.06 (0.01) & 1.15 (0.13) & 0.03 (0.01)\\ 
$100, 250, 4$ & 9.40 (0.51) & 0.12 (0.05) & 19.20 (0.49) & 0.18 (0.01) & 0.90 (0.18) & 0.05 (0.01) & 1.06 (0.13) & 0.02 (0.00)\\ 
$250, 250, 4$ & 12.00 (1.55) & 0.15 (0.06) & 19.60 (1.21) & 0.18 (0.01) & 0.52 (0.13) & 0.07 (0.01) & 1.00 (0.10) & 0.02 (0.00)\\ 
$25, 25, 10$ & 15.20 (1.80) & 0.08 (0.01) & 52.60 (3.67) & 0.06 (0.00) & 8.82 (0.33) & 0.10 (0.01) & 8.84 (0.23) & 0.08 (0.02)\\ 
$50, 25, 10$ & 14.60 (1.62) & 0.07 (0.01) & 63.00 (5.26) & 0.06 (0.01) & 7.60 (0.29) & 0.11 (0.02) & 8.48 (0.28) & 0.05 (0.01)\\ 
$100, 25, 10$ & 13.80 (1.50) & 0.06 (0.01) & 59.60 (4.80) & 0.06 (0.01) & 7.26 (0.40) & 0.10 (0.01) & 7.22 (0.28) & 0.05 (0.01)\\ 
$250, 25, 10$ & 12.80 (1.40) & 0.05 (0.01) & 49.80 (3.58) & 0.03 (0.01) & 5.70 (0.35) & 0.07 (0.01) & 6.28 (0.26) & 0.04 (0.01)\\ 
$25, 50, 10$ & 15.60 (1.75) & 0.06 (0.01) & 79.60 (2.96) & 0.06 (0.01) & 7.42 (0.35) & 0.08 (0.01) & 6.64 (0.34) & 0.07 (0.01)\\ 
$50, 50, 10$ & 14.80 (1.60) & 0.05 (0.01) & 85.80 (3.01) & 0.05 (0.01) & 6.00 (0.36) & 0.06 (0.01) & 6.12 (0.26) & 0.07 (0.01)\\ 
$100, 50, 10$ & 13.80 (1.45) & 0.04 (0.01) & 88.80 (4.35) & 0.01 (0.00) & 4.96 (0.32) & 0.05 (0.01) & 5.28 (0.32) & 0.08 (0.01)\\ 
$250, 50, 10$ & 12.60 (1.35) & 0.04 (0.01) & 89.60 (4.27) & 0.05 (0.01) & 4.36 (0.34) & 0.04 (0.01) & 4.20 (0.24) & 0.09 (0.02)\\ 
$25, 100, 10$ & 15.40 (1.70) & 0.05 (0.01) & 92.00 (6.55) & 0.03 (0.01) & 6.42 (0.28) & 0.07 (0.01) & 5.66 (0.24) & 0.04 (0.01)\\ 
$50, 100, 10$ & 14.50 (1.55) & 0.04 (0.01) & 102.60 (8.80) & 0.03 (0.01) & 5.16 (0.38) & 0.07 (0.01) & 4.98 (0.26) & 0.05 (0.01)\\ 
$100, 100, 10$ & 13.50 (1.40) & 0.03 (0.01) & 104.60 (10.10) & 0.02 (0.00) & 4.10 (0.31) & 0.06 (0.01) & 4.16 (0.28) & 0.05 (0.01)\\ 
$250, 100, 10$ & 12.00 (1.30) & 0.03 (0.01) & 99.60 (7.51) & 0.01 (0.00) & 2.82 (0.27) & 0.06 (0.01) & 3.40 (0.26) & 0.05 (0.01)\\ 
$25, 250, 10$ & 15.20 (1.65) & 0.04 (0.01) & 92.20 (3.56) & 0.02 (0.01) & 5.68 (0.30) & 0.06 (0.01) & 4.52 (0.27) & 0.04 (0.01)\\ 
$50, 250, 10$ & 14.00 (1.50) & 0.03 (0.01) & 102.20 (6.22) & 0.02 (0.01) & 3.82 (0.29) & 0.06 (0.01) & 4.23 (0.29) & 0.03 (0.01)\\ 
$100, 250, 10$ & 12.80 (1.35) & 0.03 (0.01) & 113.00 (5.94) & 0.01 (0.01) & 3.50 (0.35) & 0.05 (0.01) & 3.50 (0.21) & 0.03 (0.01)\\ 
$250, 250, 10$ & 11.50 (1.25) & 0.03 (0.01) & 110.20 (5.12) & 0.01 (0.01) & 2.30 (0.29) & 0.04 (0.01) & 3.05 (0.24) & 0.02 (0.01)\\ 
        \midrule
        \bottomrule
     \end{tabular}
  \end{threeparttable}}
\end{table}

\begin{figure}[H]
\centering
\includegraphics[width=0.5\textwidth]{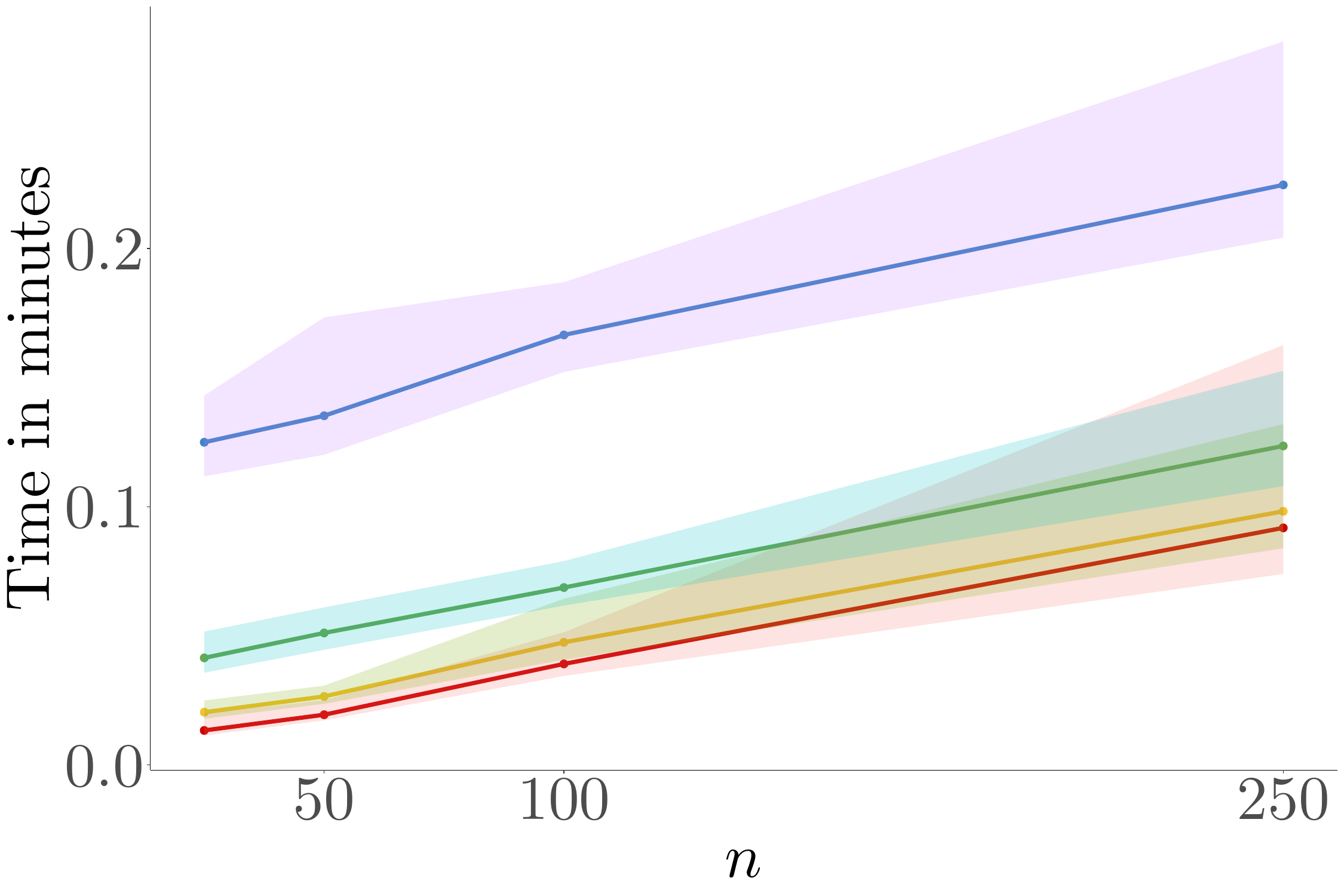}
\caption[]{Computation time for the proposed method in minutes. The colours represent the values of $m$: 25 \begin{tikzpicture}\draw [thick, colorSFROLBIC] (0,0.5) -- (0.5,0.5);\end{tikzpicture}, 50 \begin{tikzpicture}\draw [thick, colorROL] (0,0.5) -- (0.5,0.5);\end{tikzpicture}, 100 \begin{tikzpicture}\draw [thick, colorSFROLAIC] (0,0.5) -- (0.5,0.5);\end{tikzpicture} and 250 \begin{tikzpicture}\draw [thick, colorPROL] (0,0.5) -- (0.5,0.5);\end{tikzpicture}. The solid lines reflect the average times across 50 fitted models, whereas the shading around the lines reflect the minimum and maximum computation times across the fitted models.}
\label{fig:timeplot}
\end{figure}

\noindent From the results provided in Table \ref{tab:simres1}, it appears that the proposed structure learning method is capable of learning both the causal structure and the functions, provided that enough data is supplied to the model. This observation only seems to hold for the HSCM, however, as the CAM appears incapable of learning hierarchical causal structures. The estimation error for the causal functions appears to be relatively small compared to the SHD. This is likely because the main problem is within the learning of the structure itself, and it is relatively simple to learn the causal functions, as this amounts to a regression problem. For scale, it should be noted that the simulated data often takes values in $[-10,10]$, indicating that the RMSE is also small relative to the values of the data.
\\
\\
The computation time is evaluated on 50 datasets where $p = q = 10$ and $R = 2$, with different values for $n$ and $m$. The results are shown in Figure \ref{fig:timeplot}.
 
\noindent As shown in Figure \ref{fig:timeplot}, fitting the HSCM (Algorithm \ref{alg:alg_structurelearn}) is computationally inexpensive, taking well under a minute even for the largest simulated setting ($m = n = 250$, $p = q = 10$). Computing interventions (Algorithm \ref{alg:alg_intervention}), however, involves a separate and typically larger computational cost, driven primarily by the unit-level sampling step, which produces $mBN$ observations, compared with $mB$ observations at the group level. Since $B$ and $N$ are simulation parameters chosen by the user, independently of the observed sample sizes $m$ and $n$, the cost of computing an intervention can be controlled directly, but at the expense of precision. It is useful to distinguish two sources of error in the resulting interventional distribution. The first is estimation error in the causal structure and functions, inherited from the structure-learning step; this error is fixed for a given fitted HSCM and is not reduced by increasing $B$ or $N$. The second is Monte Carlo sampling error arising from the simulation itself. Because Algorithm \ref{alg:alg_intervention} uses a nested resampling scheme, in which the $N$ unit-level replicates drawn for a given group-level replicate $b$ all share the same simulated group-level ancestors, increasing $N$ alone, with $B$ fixed, primarily refines the resolution of the estimated distribution conditional on a given group-level replicate, rather than reducing the overall Monte Carlo error contributed by group-level resampling; both $B$ and $N$ generally need to increase for estimated interventional quantities to stabilize. In practice, we recommend choosing $B$ and $N$ by monitoring the stability of summary statistics of interest (e.g., the mean or selected quantiles of the interventional distribution for $X$) as $B$ and $N$ increase, and selecting the smallest values beyond which these summaries no longer change appreciably. We also note that resampling of parentless variables draws from the finite empirical distribution of the observed data (of size $m$ at the group level, or $n_j$ within group $j$ at the unit level); consequently, even as $B, N \to \infty$, the resampled marginal distributions for parentless variables converge to the empirical distribution of the observed data rather than to the (unknown) true population distribution, so the number of groups and units per group available in the original data, not just the choice of $B$ and $N$, ultimately bounds the achievable accuracy for these variables.
\section{Application on winter wheat data} \label{Application on winter wheat data}

We illustrate the proposed method using an agricultural dataset pertaining to winter wheat in Switzerland (Roth et al., \citeyear{roth2025fip}). The dataset consists of measurements across plots containing plants, which are nested within different years. Therefore, the plots reflect the units whilst the years correspond to the groups. These groups can cause heterogeneity in the growing environment for the plants, as they can correspond to different (unmeasured) weather and soil properties.  

\begin{figure}[H]
\centering
    \includegraphics[width=0.5\linewidth]{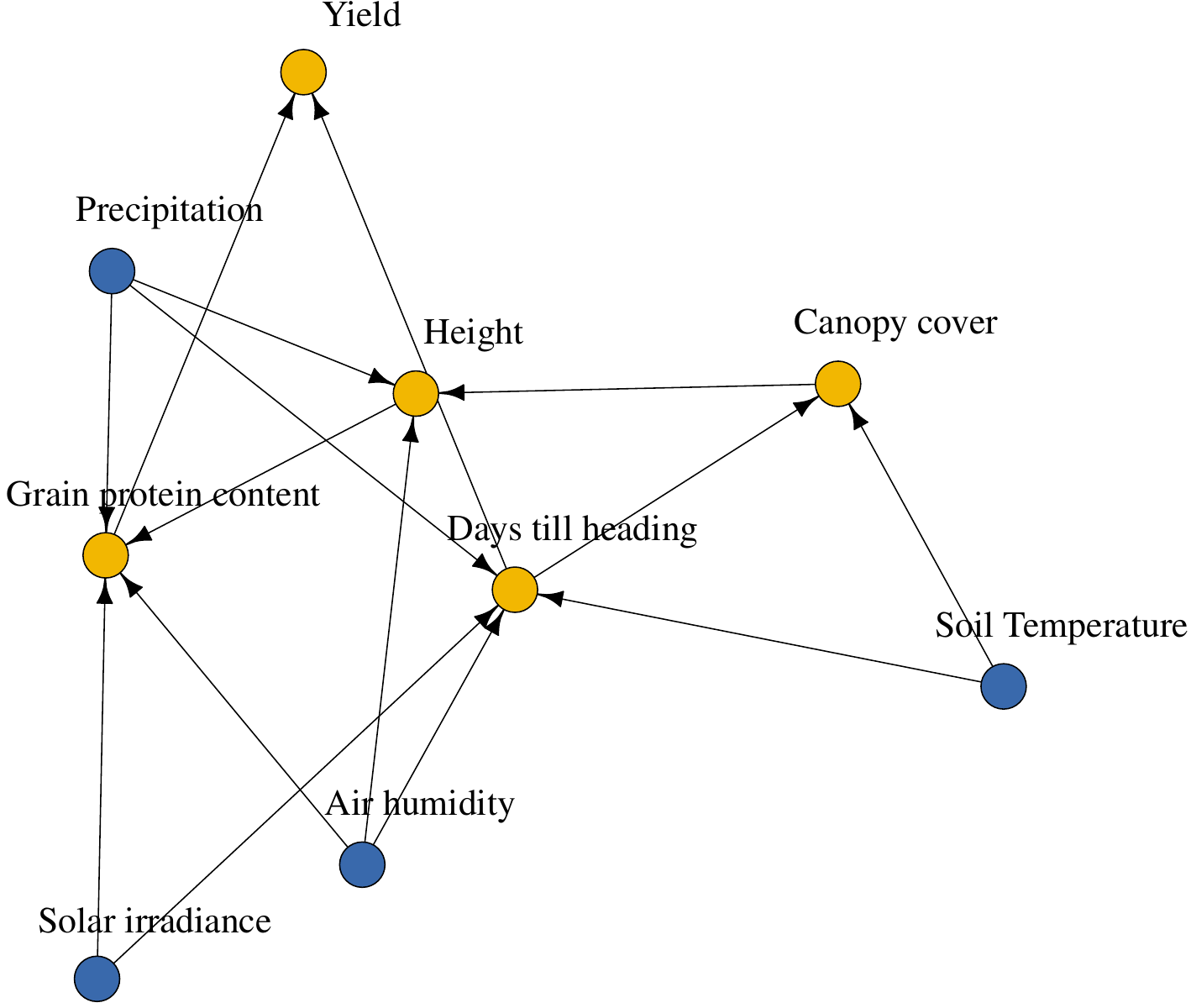}
\caption{Estimated DAG. Unit-level variables are represented by yellow vertices, whilst group-level variables are represented by blue ones} \label{fig:winter_wheat_graph}
\end{figure}

Although the dataset contains numerous variables, the central variable of interest is yield, measured at the unit level. Yield is a complex plant trait typically modelled as a function of genotype (G), environment (E), and management (M) (Hermes et al., \citeyear{hermes2023using}, \citeyear{hermes2024copula}). Its central role in agricultural research is unsurprising, as yield is a primary determinant of global food security (Fischer et al., \citeyear{fischer2014crop}). In addition to yield, the dataset includes variables characterising each of the G, E, and M components, such as plant height (G), precipitation (E), and seed rate (M). In total, the data contain 10 variables. The dataset spans four years, with group sizes ranging from 305 to 634 units (mean 459). However, the number of groups for the winter wheat data is too small to reliably estimate the group-level DAG $D_Z$. We therefore refrain from estimating  $D_Z$ and instead estimate only the relations $\bm{Z} \rightarrow \bm{X}$ and $D_X$ to avoid making potentially incorrect claims about $D_Z$.
\\
\\
We estimate the proposed method on the data, allowing for group-specific unit-level functions. The resulting causal structure is shown in Figure \ref{fig:winter_wheat_graph}.

\begin{figure}[H]
  \begin{subfigure}{0.49\textwidth}
    \includegraphics[width=\linewidth]{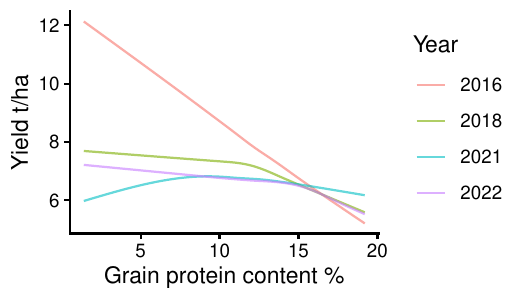}
    \caption{Estimated effect of grain protein content on yield} \label{fig:winter_wheat_protein}
  \end{subfigure}%
  \hspace*{\fill}   
  \begin{subfigure}{0.49\textwidth}
    \includegraphics[width=\linewidth]{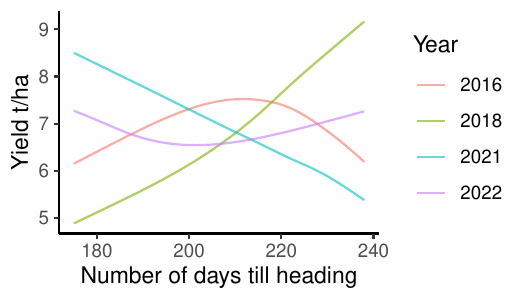}
    \caption{Estimated effect of number of days till heading on yield} \label{fig:winter_wheat_days}
  \end{subfigure}%
\caption{Estimated group specific effects for the winter wheat yield.} \label{fig:ww_rels}
\end{figure}

\noindent In addition to the causal structure, the proposed method can learn the corresponding causal functions. For both datasets, we assumed group-specific functions for the unit-unit relationships. The group-specific relationships that cause yield for winter wheat are shown in Figures \ref{fig:winter_wheat_protein} and \ref{fig:winter_wheat_days}. Scatterplots for these variables are provided in the Appendix. The causal relationships that are not directly related to yield are provided in the Appendix.

\begin{figure}[H]
  \begin{center} 
    \begin{subfigure}{0.49\textwidth}
      \includegraphics[width=\linewidth]{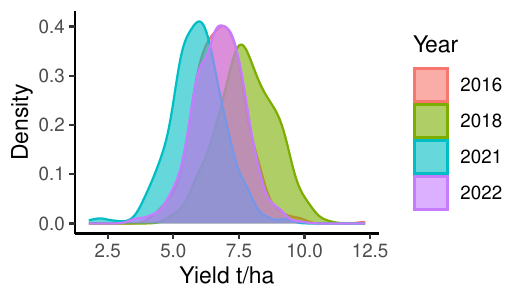}
      \caption{No intervention} \label{fig:winter_wheat_pre_intervention}
    \end{subfigure}
  \end{center}
  \begin{subfigure}{0.49\textwidth}
    \includegraphics[width=\linewidth]{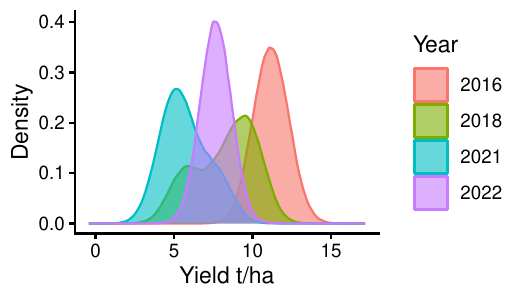}
    \caption{Intervention on grain protein content (1.4\%)} \label{fig:winter_wheat_protein_intervention_low}
  \end{subfigure}
  \hspace*{\fill}   
  \begin{subfigure}{0.49\textwidth}
    \includegraphics[width=\linewidth]{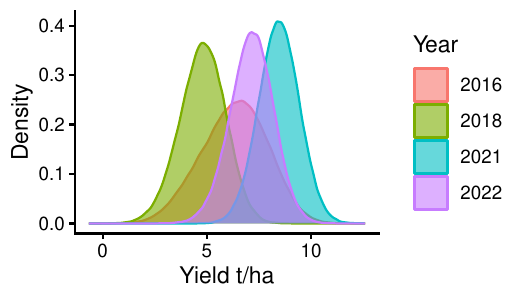}
    \caption{Intervention on number of days till heading (175 days)} \label{fig:winter_wheat_days_intervention_low}
  \end{subfigure}
  \begin{subfigure}{0.49\textwidth}
    \includegraphics[width=\linewidth]{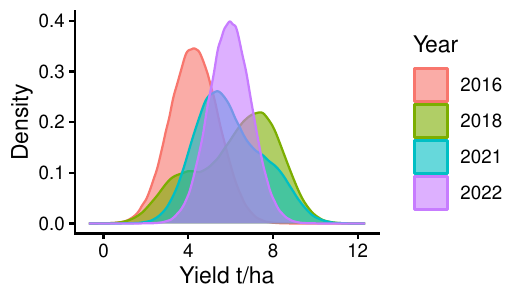}
    \caption{Intervention on grain protein content (19.2\%)} \label{fig:winter_wheat_protein_intervention_high}
  \end{subfigure}
  \hspace*{\fill}   
  \begin{subfigure}{0.49\textwidth}
    \includegraphics[width=\linewidth]{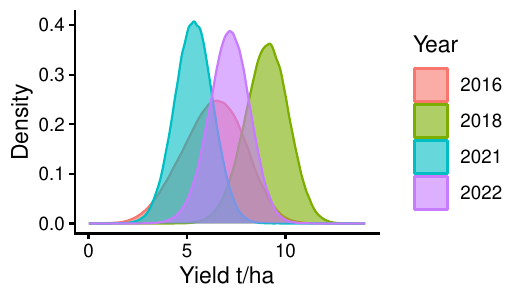}
    \caption{Intervention on number of days till heading (238 days)} \label{fig:winter_wheat_days_intervention_high}
  \end{subfigure}
\caption{Estimated interventional distributions of winter wheat yield.} \label{fig:1}
\end{figure}

\noindent The causal structure in Figure \ref{fig:winter_wheat_graph} provide us with some interesting results. As expected, yield is caused by several other variables, either directly or indirectly. Of the two direct causal relations (Figures \ref{fig:winter_wheat_protein} and \ref{fig:winter_wheat_days}), the negative relationship between grain protein content has been reported for wheat (Terman, \citeyear{terman1979yields}; Simmonds \citeyear{simmonds1995relation}; Oury \& Godin, \citeyear{oury2007yield}; Bogard et al., \citeyear{bogard2010deviation}), which may be explained by allocation of protein to photosynthesic tissues rather than grain. A causal relation with heading date is also expected, since it determines the time available for assimilation and the inconsistent shape of the relationship across years may reflect the temporal effects of weather conditions and stresses on the assimilation process. The scientific literature on this relationship is equally variable, reporting both positive (Shenoda et al., \citeyear{shenoda2021effect}; Gulino et al., \citeyear{gulino2023impact}) and negative (Riaz-ud-Din et al., \citeyear{riaz2010effect}) relations between the two variables. Most of the indirect relationships, see Appendix, also seem to make physiological sense. Days to heading will lead to more time for photosynthesis, leading to higher canopy density which would increase total assimilation and consequently higher plant biomass and height. A negative causal effect of plant height and grain protein could reflect larger allocation of protein to vegetative tissue for photosynthesis and is observed in two of the years. The deviating relation between these two traits in 2016 and 2022 is harder to explain, however.
\\
\\
As mentioned in Section \ref{Interventions}, it is possible to compute the interventional distribution for any hard intervention on a HSCM using Algorithm \ref{alg:alg_intervention}. We illustrate such interventions for the data, where we intervene on the grain protein content and the number of days till heading, by setting these to the minimum and maximum observed values for these variables (1.4\% and 19.2\% for grain protein content and 175 days and 238 days for number of days till heading).  Algorithm \ref{alg:alg_intervention} is applied on the estimated DAG and the corresponding causal functions for the data, where we set $B = N = 500$ for all interventions. The resulting interventional distributions are shown in Figures \ref{fig:winter_wheat_protein_intervention_low}, \ref{fig:winter_wheat_days_intervention_low}, \ref{fig:winter_wheat_protein_intervention_high} and \ref{fig:winter_wheat_days_intervention_high}. Moreover, we have added the pre-intervention distributions for yield throughout the years in Figure \ref{fig:winter_wheat_pre_intervention}.

\section{Conclusion} \label{Conclusion}

This paper develops a causal structure learning approach for hierarchical data in which units are nested within groups. The proposed HSCM separates causal relations into three parts: relations among group-level variables, relations among unit-level variables, and relations from group-level to unit-level variables. This separation is important because these parts of the causal structure are informed by different sources of variation in the data.

A key result is that the shared unit-level DAG $D_X$ can be learned from within-group data. Once a group is fixed, its observed and unobserved group-level contributions are constant across units and can be absorbed into the intercepts of the unit-level equations. The remaining system is a Gaussian additive noise model, allowing CAM to be used to estimate the unit-level causal structure. This avoids mixing within-group and between-group variation when learning relations between unit-level variables.

The group-to-unit relations require information across groups. After the unit-level parent sets have been determined, the observed group-level effects can be estimated jointly with the unit-level causal functions. Under the conditional mean-zero assumption for the unobserved group effects, the direct effects of the observed group-level variables can be separated from unobserved between-group variation. 

The framework also allows causal functions between unit-level variables to differ across groups while maintaining a shared unit-level DAG. A common causal function and group-specific deviations are defined on the support shared across groups, using a common centering convention. This makes it possible to describe both the causal structure that is shared between groups and differences in the shapes of the corresponding causal relations.

Because the fitted HSCM consists of causal equations, it can also be used to simulate hard interventions. This provides a way for scientists to study the consequences of changes to variables in the fitted hierarchical system.

The simulation study and winter-wheat application illustrate how the proposed framework can be used to learn and interpret causal structure in hierarchical data. More generally, the results show that distinguishing within-group from between-group information provides a useful route for extending causal discovery methods such as CAM to nested data.

Several extensions remain important. First, the present approach assumes causal sufficiency among the observed unit-level variables. Allowing unobserved variables that vary between units would broaden the range of hierarchical systems that can be studied. Second, the unit-level DAG is assumed to be shared across groups; future work could allow edges themselves, rather than only their causal functions, to differ between groups. Other useful extensions include non-Gaussian noise distributions and mixed discrete and continuous variables. Finally, the current group-specific parameterization describes the observed groups. Interventions for new or unseen groups would require an additional model for how unobserved group effects and group-specific causal functions vary across groups. Developing such a model provides a natural direction for future research.

\appendix
\section*{Appendix A: Proofs of the propositions} \label{app:proofs}
\subsection*{Proof of Proposition \ref{prop:within_group}}\label{app:proof_within_group}
\begin{proof}
Fix $j$. The values $\bm Z_j$ and $\bm A_j$ are constant with respect to the unit index $i$, and the group-specific mechanisms $s_{kd}^{(j)}$ are fixed functions within that group. Therefore the intercept, observed group-level contribution, and additive group contribution in Equation (\ref{eq:HSCM_detailed}) combine into an equation-specific constant. The remaining variable terms are precisely the functions $s_{kd}^{(j)}(X_{ij}^d)$ of the observed unit-level parents in $D_X$. By Assumption \ref{ass:sampling}, the conditional noise vectors are independent across units, their components are mutually independent across equations, and $\varepsilon_{ij}^k\sim\mathcal N(0,\sigma_{X,k}^2)$. Hence the conditional system is a Gaussian additive noise model with graph $D_X$. When Assumption \ref{ass:cam_regular} holds for this conditional system, the existing CAM identifiability result identifies $D_X$ from its observational distribution.
\end{proof}

\subsection*{Proof of Proposition \ref{prop:cross_level}}\label{app:proof_cross_level}
\begin{proof}
Because $S_k$ contains every observed group-level parent of $X^k$, substituting the group-specific form of Equation (\ref{eq:HSCM_detailed}) into Equation (\ref{eq:crosslevel_residual}) gives
\[
Q_{ij}^k
=\mu_{X,k}+\bm\beta_{k,S_k}^{\top}\bm Z_{j,S_k}
 +A_j^k+\varepsilon_{ij}^k.
\]
Assumption \ref{ass:group_conditional_mean_zero} and iterated expectation imply
\[
E(A_j^k\mid\bm Z_{j,S_k})
=E\{E(A_j^k\mid\bm Z_j)\mid\bm Z_{j,S_k}\}=0.
\]
Likewise, Equation (\ref{eq:unit_noise_mean_zero}) and iterated expectation imply
\[
E(\varepsilon_{ij}^k\mid\bm Z_{j,S_k})
=E\{E(\varepsilon_{ij}^k\mid\bm Z_j,\bm A_j)\mid\bm Z_{j,S_k}\}=0.
\]
This proves Equation (\ref{eq:crosslevel_moment}). Assumption \ref{ass:group_sampling} supplies the group-level population with respect to which the moment is defined. If two parameter vectors produce the same conditional mean almost surely, their difference $a$ satisfies $a^\top W_{j,k}=0$ almost surely. Hence
$a^\top E(W_{j,k}W_{j,k}^\top)a=0$; nonsingularity of this second-moment matrix gives $a=0$.
\end{proof}

\section*{Appendix B: Additional simulation results} \label{Additional simulation results}

In addition to the simulations for the baseline model, that is the model presented by Equation (\ref{eq:HSCM_detailed}), causal structure learning for the proposed extensions is shown by means of additional simulations. The data generating mechanism is exactly the same as that for the simulations in Section \ref{Simulation study}, except that all errors are drawn from $U\left(-1,1\right)$. The results are presented in Table \ref{tab:simres3}

\begin{table}[H]
\centering
\scalebox{1}{%
  \begin{threeparttable}
  \caption{Results for the proposed method under misspecified settings. The SHD and RMSE are averaged across 50 randomly generated datasets.}
  \label{tab:simres3}
     \begin{tabular}{c | cc}
        \toprule
        \midrule
         $\bm{n, m, p, q}$ & \textbf{SHD} & \textbf{RMSE}\\ \midrule
$25, 25, 4, 4$ & 3.84 (0.22) & 0.17 (0.03)\\ 
$50, 25, 4, 4$ & 3.74 (0.23) & 0.15 (0.03)\\  
$100, 25, 4, 4$ & 3.72 (0.20) & 0.15 (0.02)\\ 
$250, 25, 4, 4$ & 3.42 (0.21) & 0.16 (0.03)\\ 
 $25, 50, 4, 4$ & 2.98 (0.16) & 0.11 (0.01)\\  
$50, 50, 4, 4$ & 2.84 (0.21) & 0.10 (0.01)\\ 
$100, 50, 4, 4$ & 2.58 (0.19) & 0.09 (0.01)\\  
$250, 50, 4, 4$ & 2.78 (0.24) & 0.11 (0.02)\\  
$25, 100, 4, 4$ & 2.98 (0.21) & 0.12 (0.02)\\ 
$50, 100, 4, 4$ & 2.70 (0.23) & 0.11 (0.02)\\ 
$100, 100, 4, 4$ & 3.06 (0.26) & 0.13 (0.03)\\ 
$250, 100, 4, 4$ & 2.78 (0.28) & 0.12 (0.03)\\ 
$25, 250, 4, 4$ & 2.58 (0.18) & 0.08 (0.01)\\ 
$50, 250, 4, 4$ & 1.90 (0.20) & 0.07 (0.01)\\ 
$100, 250, 4, 4$ & 1.80 (0.18) & 0.08 (0.02)\\ 
$250, 250, 4, 4$ & 2.10 (0.20) & 0.08 (0.01)\\ 
$25, 25, 10, 10$ & 9.78 (0.28) & 0.12 (0.01)\\ 
$50, 25, 10, 10$ & 8.86 (0.32) & 0.10 (0.01)\\ 
$100, 25, 10, 10$ & 8.88 (0.46) & 0.10 (0.01)\\  
$250, 25, 10, 10$ & 8.22 (0.38) & 0.09 (0.01)\\ 
$25, 50, 10, 10$ & 9.14 (0.35) & 0.11 (0.01)\\ 
$50, 50, 10, 10$ & 8.10 (0.36) & 0.11 (0.01)\\ 
$100, 50, 10, 10$ & 7.16 (0.41) & 0.10 (0.01)\\ 
$250, 50, 10, 10$ & 6.88 (0.36) & 0.09 (0.01)\\ 
$25, 100, 10, 10$ & 8.52 (0.44) & 0.10 (0.01)\\ 
$50, 100, 10, 10$ & 6.40 (0.47) & 0.10 (0.01)\\ 
$100, 100, 10, 10$ & 6.52 (0.51) & 0.10 (0.01)\\  
$250, 100, 10, 10$ & 6.34 (0.46) & 0.10 (0.01)\\ 
$25, 250, 10, 10$ & 7.56 (0.29) & 0.08 (0.01)\\  
$50, 250, 10, 10$ & 6.16 (0.34) & 0.08 (0.01)\\ 
$100, 250, 10, 10$ & 5.30 (0.43) & 0.08 (0.01)\\ 
$250, 250, 10, 10$ & 5.16 (0.47) & 0.08 (0.01)\\ 
        \midrule
        \bottomrule
     \end{tabular}
  \end{threeparttable}}
\end{table}

\section*{Appendix C: Additional winter wheat results} \label{Additional winter wheat results}

\begin{figure}[H]
  \begin{subfigure}{0.49\textwidth}
    \includegraphics[width=\linewidth]{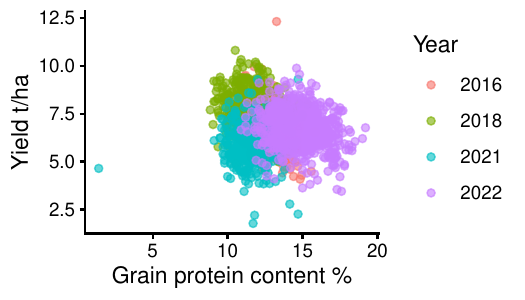}
    \caption{Scatterplot between grain protein content and yield} \label{fig:FIP_yield_protein_scatter}
  \end{subfigure}%
  \hspace*{\fill}   
  \begin{subfigure}{0.49\textwidth}
    \includegraphics[width=\linewidth]{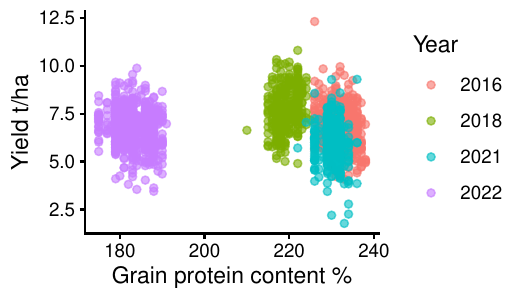}
    \caption{Scatterplot between number of days till heading and yield} \label{fig:FIP_yield_days_till_heading_scatter}
  \end{subfigure}%
\caption{Scatterplot between winter wheat yield and its causes.} \label{fig:ww_rels}
\end{figure}

Even though Figures \ref{fig:winter_wheat_protein} and \ref{fig:winter_wheat_days} provide the estimated causal functions for yield, some readers might be interested in the raw bivariate relationships found in the data. We illustrate this by means of scatterplots, which are presented in Figures \ref{fig:FIP_yield_protein_scatter} and \ref{fig:FIP_yield_days_till_heading_scatter}.

\begin{figure}[H]
  \begin{center} 
    \begin{subfigure}{0.49\textwidth}
      \includegraphics[width=\linewidth]{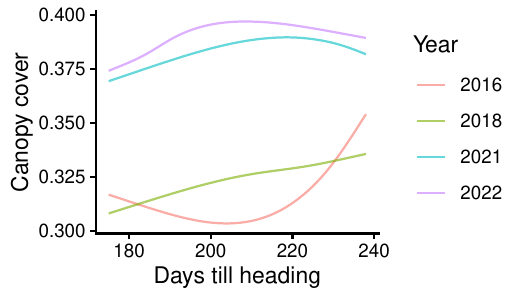}
      \caption{Estimated effect of days till heading on canopy cover} \label{fig:winter_wheat_heading_canopy}
    \end{subfigure}
  \end{center}
  \begin{subfigure}{0.49\textwidth}
    \includegraphics[width=\linewidth]{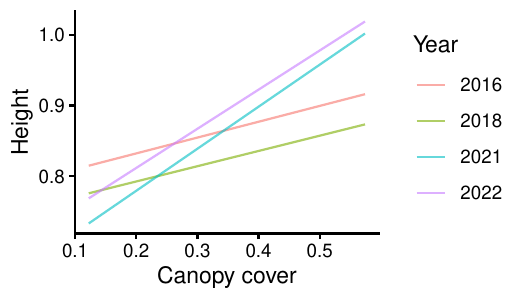}
    \caption{Estimated effect of canopy cover on height} \label{fig:winter_wheat_canopy_height}
  \end{subfigure}
  \hspace*{\fill}   
  \begin{subfigure}{0.49\textwidth}
    \includegraphics[width=\linewidth]{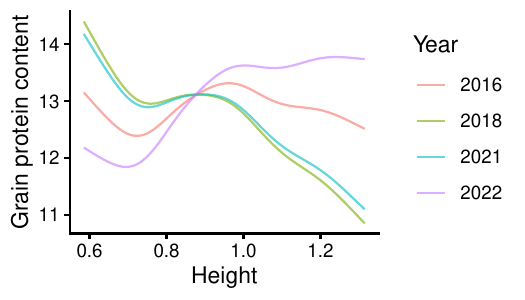}
    \caption{Estimated effect of height on protein content} \label{fig:winter_wheat_height_protein}
      \end{subfigure}
\caption{Estimated non-yield group specific effects for winter wheat} \label{fig:ex1}
\end{figure}

\noindent In addition to the results presented in Section \ref{Application on winter wheat data}, the HSCM provided results between the non-yield variables. These are presented in Figures \ref{fig:winter_wheat_heading_canopy}, \ref{fig:winter_wheat_canopy_height} and \ref{fig:winter_wheat_height_protein}.

\bibliographystyle{Chicago}
\bibliography{library}

@inproceedings{witty2020causal,
  title={Causal inference using Gaussian processes with structured latent confounders},
  author={Witty, Sam and Takatsu, Kenta and Jensen, David and Mansinghka, Vikash},
  booktitle={International Conference on Machine Learning},
  pages={10313--10323},
  year={2020},
  organization={PMLR}
}

@article{weinstein2024hierarchical,
  title={Hierarchical causal models},
  author={Weinstein, Eli N and Blei, David M},
  journal={Journal of Machine Learning Research},
  volume={27},
  number={37},
  pages={1--73},
  year={2026}
}

@book{peters2017elements,
  title={Elements of causal inference: foundations and learning algorithms},
  author={Peters, Jonas and Janzing, Dominik and Sch{\"o}lkopf, Bernhard},
  year={2017},
  publisher={The MIT Press}
}

@article{peters2014causal,
  title={Causal discovery with continuous additive noise models},
  author={Peters, Jonas and Mooij, Joris M and Janzing, Dominik and Sch{\"o}lkopf, Bernhard},
  journal={The Journal of Machine Learning Research},
  volume={15},
  number={1},
  pages={2009--2053},
  year={2014},
  publisher={JMLR. org}
}

@book{pearl2009causality,
  title={Causality},
  author={Pearl, Judea},
  year={2009},
  publisher={Cambridge university press}
}

@inproceedings{jensen2020object,
  title={Object conditioning for causal inference},
  author={Jensen, David and Burroni, Javier and Rattigan, Matthew},
  booktitle={Uncertainty in Artificial Intelligence},
  pages={1072--1082},
  year={2020},
  organization={PMLR}
}

@article{hermes2024copula,
  title={Copula graphical models for heterogeneous mixed data},
  author={Hermes, Sjoerd and van Heerwaarden, Joost and Behrouzi, Pariya},
  journal={Journal of Computational and Graphical Statistics},
  volume={33},
  number={3},
  pages={991--1005},
  year={2024},
  publisher={Taylor \& Francis}
}

@article{van2005statistical,
  title={Statistical models for genotype by environment data: from conventional ANOVA models to eco-physiological QTL models},
  author={van Eeuwijk, Fred A and Malosetti, Marcos and Yin, Xinyou and Struik, Paul C and Stam, Piet},
  journal={Australian Journal of Agricultural Research},
  volume={56},
  number={9},
  pages={883--894},
  year={2005},
  publisher={CSIRO Publishing}
}

@article{buhlmann2014cam,
  title={CAM: Causal additive models, high-dimensional order search and penalized regression},
  author={B{\"u}hlmann, Peter and Peters, Jonas and Ernest, Jan},
  journal={The Annals of Statistics},
  pages={2526--2556},
  year={2014},
  publisher={JSTOR}
}

@article{nowzohour2016score,
  title={Score-based causal learning in additive noise models},
  author={Nowzohour, Christopher and B{\"u}hlmann, Peter},
  journal={Statistics},
  volume={50},
  number={3},
  pages={471--485},
  year={2016},
  publisher={Taylor \& Francis}
}

@article{roth2025fip,
  title={The FIP 1.0 Data Set: Highly resolved annotated image time series of 4,000 wheat plots grown in 6 years},
  author={Roth, Lukas and Boss, Mike and Kirchgessner, Norbert and Aasen, Helge and Aguirre-Cuellar, Brenda Patricia and Akiina, Price Pius Atuah and Anderegg, Jonas and Castillo, Joaquin Gajardo and Chen, Xiaoran and Corrado, Simon and others},
  journal={GigaScience},
  volume={14},
  pages={giaf051},
  year={2025},
  publisher={Oxford University Press}
}

@article{fischer2014crop,
  title={Crop yields and global food security},
  author={Fischer, RA and Byerlee, Derek and Edmeades, Greg},
  journal={ACIAR: Canberra, ACT},
  pages={8--11},
  year={2014}
}

@article{hermes2023using,
  title={Using copula graphical models to detect the impact of drought stress on maize and wheat yield},
  author={Hermes, Sjoerd and van Heerwaarden, Joost and Behrouzi, Pariya},
  journal={in silico Plants},
  volume={5},
  number={1},
  pages={diad008},
  year={2023},
  publisher={Oxford University Press UK}
}

@article{terman1979yields,
  title={Yields and Protein Content of Wheat Grain as Affected by Cultivar, N, and Environmental Growth Factors 1},
  author={Terman, GL},
  journal={Agronomy Journal},
  volume={71},
  number={3},
  pages={437--440},
  year={1979},
  publisher={Wiley Online Library}
}

@article{oury2007yield,
  title={Yield and grain protein concentration in bread wheat: how to use the negative relationship between the two characters to identify favourable genotypes?},
  author={Oury, Fran{\c{c}}ois-Xavier and Godin, Christelle},
  journal={Euphytica},
  volume={157},
  number={1},
  pages={45--57},
  year={2007},
  publisher={Springer}
}

@article{bogard2010deviation,
  title={Deviation from the grain protein concentration--grain yield negative relationship is highly correlated to post-anthesis N uptake in winter wheat},
  author={Bogard, Matthieu and Allard, Vincent and Brancourt-Hulmel, Maryse and Heumez, Emmanuel and Machet, Jean-Marie and Jeuffroy, Marie-H{\'e}l{\`e}ne and Gate, Philippe and Martre, Pierre and Le Gouis, Jacques},
  journal={Journal of experimental botany},
  volume={61},
  number={15},
  pages={4303--4312},
  year={2010},
  publisher={Oxford University Press}
}

@article{shenoda2021effect,
  title={Effect of long-term heat stress on grain yield, pollen grain viability and germinability in bread wheat (Triticum aestivum L.) under field conditions},
  author={Shenoda, JE and Sanad, Marwa NME and Rizkalla, Aida A and El-Assal, S and Ali, Rania T and Hussein, Mona H},
  journal={Heliyon},
  volume={7},
  number={6},
  year={2021},
  publisher={Elsevier}
}

@article{gulino2023impact,
  title={Impact of rising temperatures on historical wheat yield, phenology, and grain size in Catalonia},
  author={Gulino, Davide and Sayeras, Roser and Serra, Joan and Betbese, Josep and Doltra, Jordi and Gracia-Romero, Adrian and Lopes, Marta S},
  journal={Frontiers in Plant Science},
  volume={14},
  pages={1245362},
  year={2023},
  publisher={Frontiers Media SA}
}

@article{riaz2010effect,
  title={Effect of temperature on development and grain formation in spring wheat},
  author={Riaz-ud-Din, MS and Ahmad, Naeem and Hussain, Makhdoom and Rehman, Aziz Ur},
  journal={Pak. J. Bot},
  volume={42},
  number={2},
  pages={899--906},
  year={2010}
}

@article{mooij2016distinguishing,
  title={Distinguishing cause from effect using observational data: methods and benchmarks},
  author={Mooij, Joris M and Peters, Jonas and Janzing, Dominik and Zscheischler, Jakob and Sch{\"o}lkopf, Bernhard},
  journal={Journal of Machine Learning Research},
  volume={17},
  number={32},
  pages={1--102},
  year={2016}
}

@inproceedings{mooij2009regression,
  title={Regression by dependence minimization and its application to causal inference in additive noise models},
  author={Mooij, Joris and Janzing, Dominik and Peters, Jonas and Sch{\"o}lkopf, Bernhard},
  booktitle={Proceedings of the 26th annual international conference on machine learning},
  pages={745--752},
  year={2009}
}

@article{simmonds1995relation,
  title={The relation between yield and protein in cereal grain},
  author={Simmonds, Norman W},
  journal={Journal of the Science of Food and Agriculture},
  volume={67},
  number={3},
  pages={309--315},
  year={1995},
  publisher={Wiley Online Library}
}

\end{document}